\documentclass[twocolumn,prb,showpacs]{revtex4}

\usepackage{graphicx}
\usepackage{dcolumn}
\usepackage{bm}
\usepackage{amssymb, amsmath}
\usepackage{color}

\begin{document}
\title{Consistent two-lifetime model for 
spectral functions of superconductors
} 

\author{Franti\v{s}ek Herman and Richard Hlubina}

\affiliation{Department of Experimental Physics, Comenius University,
  Mlynsk\'{a} Dolina F2, 842 48 Bratislava, Slovakia}

\begin{abstract}
Recently it has been found that models with at least two lifetimes
have to be considered when analyzing the angle resolved photoemission
data in the nodal region of the cuprates [T. Kondo et al.,
  Nat. Commun. 6, 7699 (2015)]. In this paper we compare two such
models. First we show that the phenomenological model used by Kondo et
al.  violates the sum rule for the occupation number.  Next we
consider the recently proposed model of the so-called Dynes
superconductors, wherein the two lifetimes measure the strengths of
pair-conserving and pair-breaking processes. We demonstrate that the
model of the Dynes superconductors is fully consistent with known
exact results and we study in detail the resulting spectral
functions. Finally, we show that the spectral functions in the nodal
region of the cuprates can be fitted well by the model of the Dynes
superconductors.

\end{abstract}
\pacs{PACS}
\maketitle

\section{Introduction}
Recent experimental progress in angle resolved photoemission
spectroscopy (ARPES)\cite{Hashimoto14} enables not only to determine
the position of features in the electron spectral function, but also
to study more subtle issues such as the spectral lineshapes. At least
for the conventional low-temperature superconductors, there does exist
a theoretical technique which can address such issues - namely the
Eliashberg theory,\cite{Marsiglio08} which allows for strong coupling
between the electrons and bosonic collective modes. However, it is not
obvious whether this type of theory is applicable to the
cuprates. Moreover, even conventional superconductors may possess
complicated phonon spectra and/or they can exhibit substantial elastic
scattering,\cite{Szabo16} both of which complicate the Eliashberg
analysis. For all these reasons, it is desirable to have a simple and
generic theory which does take finite quasiparticle lifetimes in
superconductors into account. 

Let us start by noting that spectral functions of a BCS superconductor
in presence of elastic pair-conserving scattering can be found even in
textbooks.\cite{Zhu04,Marsiglio08} However, it is well known that the
density of states (or the so-called tomographic density of states in
case of anisotropic superconductors\cite{Reber12}) implied by such
spectral functions exhibits a full spectral gap consistent with the
Anderson theorem. On the other hand, experimentally, the gap is quite
often only partial\cite{Szabo16,Reber12} and the tunneling density of
states is better described by the phenomenological Dynes
formula.\cite{Dynes78,White86} This means then that, in order to take
the non-trivial density of states into account, also a second type of
scattering processes - which are not subject to the Anderson theorem -
have to be considered. Two-lifetime phenomenology of precisely this
type has in fact been applied quite recently,\cite{Kondo15} with the
aim to parameterize the high-resolution ARPES data in the nodal region
of the cuprates.

The goals of this paper are twofold. First, in Section~2 we
demonstrate that the model used in Ref.~\onlinecite{Kondo15} can be
cast into a form consistent with the generalized Eliashberg theory,
and that it exhibits several attractive features. However, we also
show that the resulting $2\times 2$ Nambu-Gor'kov propagator violates
the sum rule for the occupation number and therefore the model used in
Ref.~\onlinecite{Kondo15} should be discarded.

Our second goal is to demonstrate that, nevertheless, a fully
consistent two-lifetime phenomenology for superconductors does exist.
To this end we consider the recently proposed model of the so-called
Dynes superconductors,\cite{Herman16} wherein the two-lifetime
phenomenology results as a consequence of taking into account both,
the pair-conserving and the pair-breaking scattering processes.  In
Section~3 we present detailed predictions for the spectral functions
of the Dynes superconductors and we explicitly demonstrate the
applicability of this approach to the low-temperature ARPES data in
the nodal region of the cuprates. Furthermore, in the Appendices we
show that the model of the Dynes superconductors is fully
consistent with known exact results, and we present explicit formulas
for the momentum distribution functions\cite{Campuzano04} within the
Eliashberg theory. Finally, in Section~4 we present our conclusions.

\section{The model used by Kondo et al.}
Following previous theoretical suggestions,\cite{Norman98,Chubukov07}
the authors of Ref.~\onlinecite{Kondo15} fit their high resolution
ARPES data in the nodal region of optimally doped and overdoped Bi2212
samples to spectral functions derived from the phenomenological self
energy
\begin{equation}
\Sigma(\mathbf{k},\omega) = -i \Gamma_1 
+ \frac{\overline{\Delta}^2}
{\omega + \varepsilon_{\mathbf{k}}+i\Gamma_0}
\label{eq:SelfEnergy}
\end{equation}
 and they interpret the scattering rates $\Gamma_1$ and $\Gamma_0$ as
 the single-particle and pair scattering rates, respectively. We shall
 comment on these identifications later.

In order to demonstrate the physical meaning of the phenomenological
self energy Eq.~\eqref{eq:SelfEnergy}, let us first note that it
implies that the electron Green function in the superconducting state
can be written in the form
\begin{equation}
G(\mathbf{k},\omega) = 
\frac{(\omega + i\gamma) + (\varepsilon_{\mathbf{k}} + i\gamma')}
{(\omega + i\gamma)^2 - (\varepsilon_{\mathbf{k}} + i\gamma')^2 
- \overline{\Delta}^2},
\label{eq:kondo_green}
\end{equation}
where we have introduced $\gamma = (\Gamma_0 + \Gamma_1)/2$ and
$\gamma' = (\Gamma_0 - \Gamma_1)/2$. According to
Ref.~\onlinecite{Kondo15}, throughout the superconducting phase
$\gamma'<0$. With increasing temperature $|\gamma'|$ decreases and it
vanishes at the critical temperature.

The main observation of this Section is that
Eq.~\eqref{eq:kondo_green} should form the upper left component of the
general $2\times 2$ Nambu-Gor'kov Green function for a superconductor:
\begin{equation}
\label{eq:G_g}
{\hat G}(\mathbf{k},\omega) = \frac{\omega Z(\mathbf{k},\omega)\tau_0 +
  \left[\varepsilon_{\mathbf{k}} + \chi(\mathbf{k},\omega)\right]\tau_3 +
  \phi(\mathbf{k},\omega)\tau_1}{\left[\omega Z(\mathbf{k},\omega)\right]^2 -
  \left[\varepsilon_{\mathbf{k}} + \chi(\mathbf{k},\omega)\right]^2 -
  \phi(\mathbf{k},\omega)^2},
\end{equation}
where $\tau_i$ are the Pauli matrices, $Z(\mathbf{k},\omega)$ is the
wave-function renormalization, and $\phi(\mathbf{k},\omega)$ is the
anomalous self-energy.  The function $\chi(\mathbf{k},\omega)$
describes the renormalization of the single-particle spectrum and it
vanishes in a particle-hole symmetric theory; that is why it is
usually neglected. For the sake of completeness, let us mention that
the matrix $\tau_2$ does not enter Eq.~\eqref{eq:G_g}, because we work
in a gauge with a real order parameter.

Since in the high-frequency limit the functions $\chi$, $\phi$, and
$\omega(Z-1)$ should stay at most constant, Eq.~\eqref{eq:kondo_green}
in this limit can be uniquely interpreted in terms of
the general expression Eq.~\eqref{eq:G_g} with
\begin{equation}
Z(\omega)=1+i\gamma/\omega,
\quad 
\chi(\omega) = i\gamma', 
\quad 
\phi(\omega) = \overline{\Delta},
\label{eq:Z}
\end{equation}
where we have chosen a real anomalous self-energy $\phi$. One checks
readily that Eq.~\eqref{eq:Z} reproduces Eq.~\eqref{eq:kondo_green}
for all frequencies. The finite value of $\chi$ is unusual but seems
to be attractive, since the cuprates, being doped Mott insulators,
might be expected to break the particle-hole symmetry. Note also that
all three functions $Z$, $\chi$ and $\phi$ do not depend on ${\bf k}$,
which is the standard behavior.

It is well known that the $2\times 2$ formalism leads to a redundant
description, and therefore the Green function has to satisfy
additional constraints. These constraints are most clearly visible in
the Matsubara formalism, therefore let us reformulate
Eq.~\eqref{eq:G_g} on the imaginary axis, allowing explicitly only for
frequency-dependent functions $Z_n=Z(i\omega_n)$,
$\chi_n=\chi(i\omega_n)$, and $\phi_n=\phi(i\omega_n)$:
\begin{equation}
{\hat G}(\mathbf{k},\omega_n) = -\frac{i\omega_n Z_n\tau_0 +
  \big(\varepsilon_{\mathbf{k}} + \chi_n\big)\tau_3 +
  \phi_n\tau_1}{\big(\omega_n Z_n\big)^2 +
  \big(\varepsilon_{\mathbf{k}} + \chi_n\big)^2 + \phi_n^2}.
\label{eq:matsubara_green}
\end{equation}
Due to the redundancy of the $2\times 2$ formalism, singlet
superconductors have to exhibit the following symmetry:
\begin{equation}
G_{22}(\mathbf{k},\omega_n)=-G_{11}(\mathbf{k},-\omega_n).
\label{eq:symmetry}
\end{equation}
Note that the functions Eq.~\eqref{eq:Z} read as
$Z_n=1+i\gamma/|\omega_n|$, $\chi_n = i\gamma'$, and $\phi_n =
\overline{\Delta}$ on the imaginary axis. It is easy to see that when
these expressions are plugged in into Eq.~\eqref{eq:matsubara_green},
the Green function does satisfy Eq.~\eqref{eq:symmetry}.

An additional attractive feature of the phenomenology Eq.~\eqref{eq:Z}
is that it leads (also for finite values of $\gamma^\prime$) to the
Dynes formula for the tunneling density of states,
\begin{equation}
N(\omega)=N_0{\rm Re}\left[\frac{\omega+i\gamma}
{\sqrt{(\omega+i\gamma)^2-{\bar \Delta}^2}}\right],
\label{eq:dynes}
\end{equation}
in agreement with the experimental findings of
Ref.~\onlinecite{Reber12}.  The square root has to be taken so that
its imaginary part is positive and we keep this convention throughout
this paper.  In Eq.~(\ref{eq:dynes}) $N_0$ denotes the normal-state
density of states. Note that, although the particle-hole symmetry is
broken due to $\chi\neq 0$, $N(\omega)$ is an even function of
$\omega$.

The broken particle-hole symmetry is clearly visible already in the
normal state with $\overline{\Delta}=0$, in which case 
\begin{equation}
G_{11}(\mathbf{k},\omega)= 
\frac{1}{\omega - \varepsilon_{\mathbf{k}} + i\Gamma_1}, 
\quad
G_{22}(\mathbf{k},\omega)= 
\frac{1}{\omega + \varepsilon_{\mathbf{k}} + i\Gamma_0}.
\label{eq:normal}
\end{equation}
These results show that $\Gamma_0$ and $\Gamma_1$ should not be
interpreted as pair and single-particle scattering rates as has been
done in Ref.~\onlinecite{Kondo15}, but rather as the scattering rates
for the holes and for the electrons, respectively.

\begin{figure}[t]
\includegraphics[width=7.0cm]{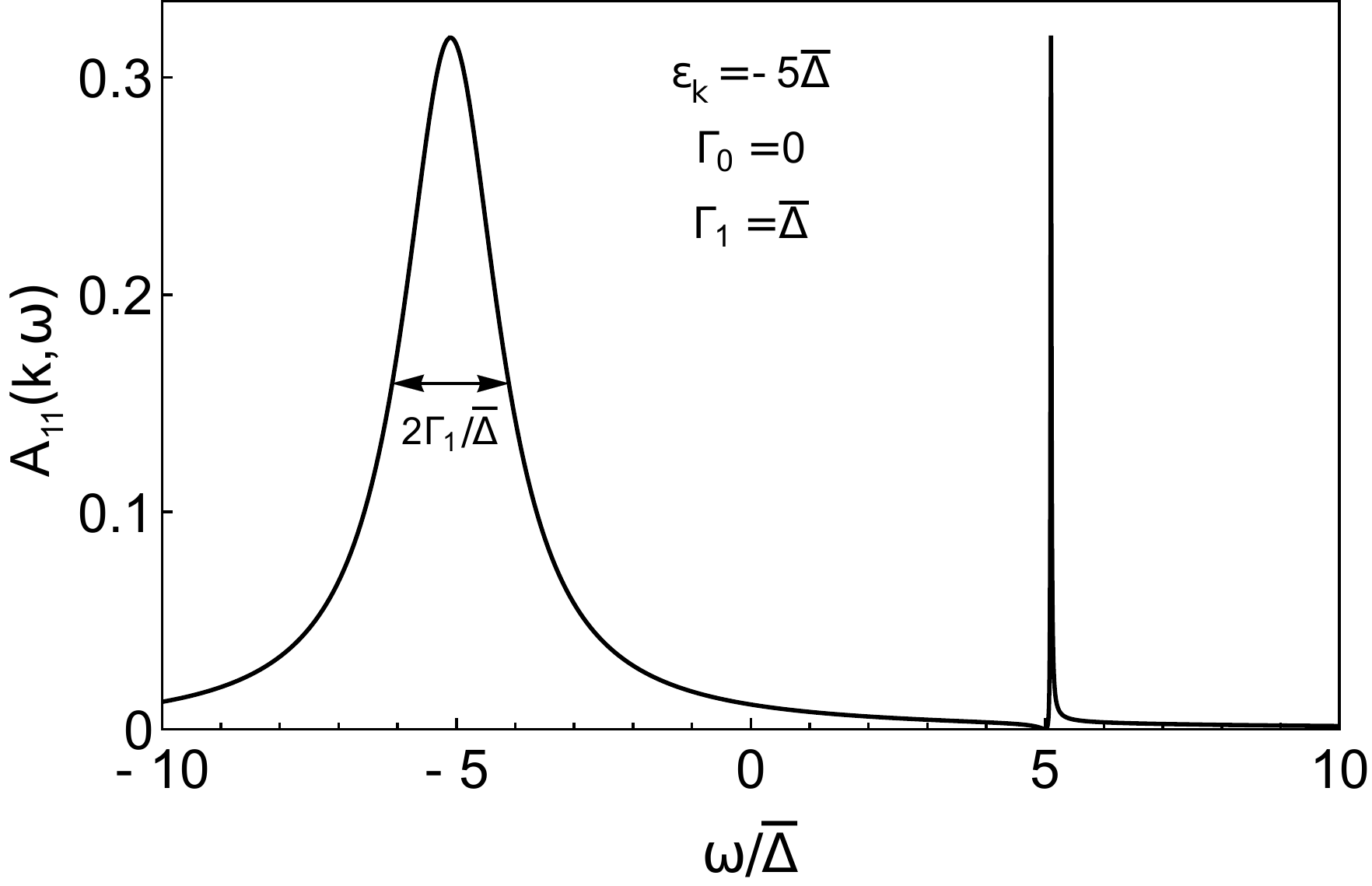}
\caption{Spectral function $A_{11}(\mathbf{k},\omega)$ of an electron
  inside the Fermi sea according to the
  model~\eqref{eq:kondo_green}. The widths of the electron-like branch
  at $\omega\approx -E_{\bf k}$ and of the hole-like branch at
  $\omega\approx E_{\bf k}$, where $E_{\bf k}$ is the quasiparticle
  energy Eq.~\eqref{eq:dispersion}, are essentially determined by
  $\Gamma_1$ and $\Gamma_0$, respectively.}
\label{fig:kondo}
\end{figure}

\begin{figure*}[t]
\includegraphics[width = 17cm]{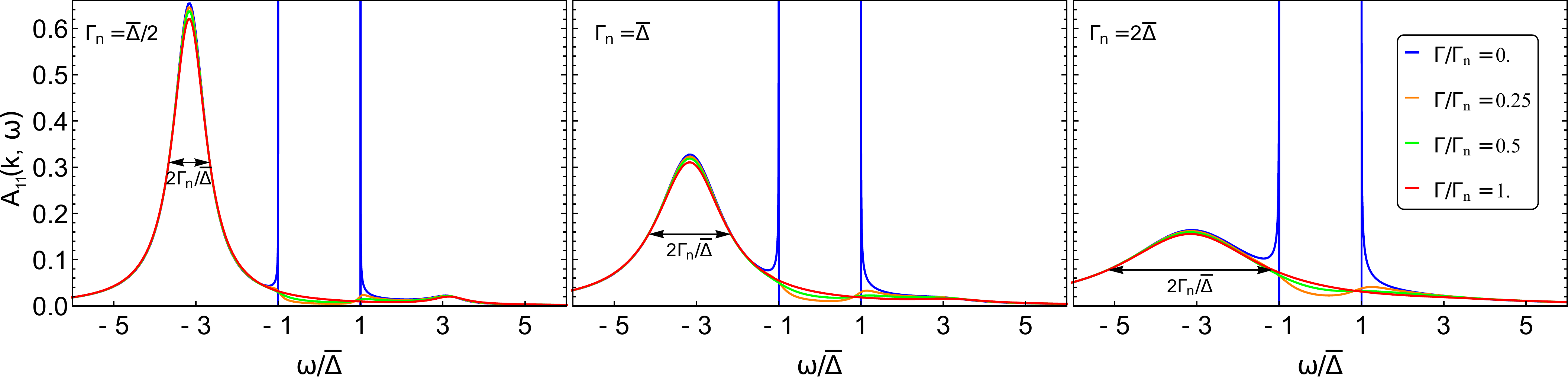}
\caption{Spectral functions $A_{11}(\mathbf{k},\omega)$ of the Dynes
  superconductor for an electron inside the Fermi sea with
  $\varepsilon_{\bf k}=-3\overline{\Delta}$. The total scattering rate
  $\Gamma_n=\Gamma+\Gamma_s$ increases from the left to the right
  panel. The curves in each panel differ by the strength of the
  pair-breaking scattering rate $\Gamma$, while $\Gamma_n$ is kept
  fixed. The color coding is the same in all panels.}
\label{fig:arpes1}
\end{figure*}

Unfortunately, Eqs.~\eqref{eq:normal} turn out to be mutually
inconsistent, but in a quite subtle way.  In order to show this, let
us introduce the spectral functions $A_{ii}(\mathbf{k},\omega)$ with
$i=1,2$, corresponding to the Green functions
$G_{ii}(\mathbf{k},\omega)$. Applying standard procedures, the
following exact sum rules can be established,
\begin{eqnarray*}
\int_{-\infty}^{\infty}\frac{d\omega}{1+e^{-\omega/T}}
A_{11}(\mathbf{k},\omega)
&=& \langle c^{}_{{\bf k}\uparrow}c^\dagger_{{\bf k}\uparrow}\rangle
=1-n^{}_{{\bf k}\uparrow},
\\
\int_{-\infty}^{\infty}\frac{d\omega}{1+e^{-\omega/T}}
A_{22}(\mathbf{k},\omega)
&=& \langle c^\dagger_{-{\bf k}\downarrow}c^{}_{-{\bf k}\downarrow}\rangle
=n^{}_{-{\bf k}\downarrow}.
\end{eqnarray*}
But in a singlet superconductor we have $n^{}_{{\bf
    k}\uparrow}=n^{}_{-{\bf k}\downarrow}$, and therefore the
following exact relation should hold:
\begin{equation}
\int_{-\infty}^{\infty}\frac{d\omega}{1+e^{-\omega/T}}
\left[A_{11}(\mathbf{k},\omega)+A_{22}(\mathbf{k},\omega)\right]=1.
\label{eq:sumrule}
\end{equation}
Making use of Eqs.~\eqref{eq:normal} at $T=0$, the integrals on the
left-hand side can be taken easily and the results are
\begin{eqnarray*}
\int_{0}^{\infty}d\omega A_{11}(\mathbf{k},\omega) &=& \frac{1}{2}
+\frac{1}{\pi}\arctan\left(\frac{\varepsilon_{\mathbf{k}}}{\Gamma_1}\right), 
\\
\int_{0}^{\infty}d\omega A_{22}(\mathbf{k},\omega) &=& \frac{1}{2}
-\frac{1}{\pi}\arctan\left(\frac{\varepsilon_{\mathbf{k}}}{\Gamma_0}\right).
\end{eqnarray*}
It can be seen readily that, if $\Gamma_0\neq \Gamma_1$, the sum
rule~\eqref{eq:sumrule} is violated by these results.

One might have the impression that the particle-hole asymmetry which
causes the sum rule violation is an artifact of our generalization of
the Green function~\eqref{eq:kondo_green} to the matrix
form~\eqref{eq:G_g}. That this is not the case can be seen by plotting
the spectral function directly for Eq.~\eqref{eq:kondo_green}, see
Fig.~\ref{fig:kondo}, which clearly shows that the electron- and
hole-like branches exhibit different scattering rates.

We conclude that the phenomenology~\eqref{eq:Z} is internally
consistent only if $\Gamma_0=\Gamma_1=\gamma$, in which case
$\gamma^\prime=0$. But then the Green function~\eqref{eq:G_g} has a
simple two-pole structure with poles at $\omega=\pm E_{\bf
  k}-i\gamma$ where
\begin{equation}
E_{\bf k}=\sqrt{\varepsilon_{\bf k}^2+{\overline{\Delta}}^2},
\label{eq:dispersion}
\end{equation}
implying that the spectral function is a sum of two Lorentzians.
However, the authors of Ref.~\onlinecite{Kondo15} stress that the
experimentally observed lineshapes are asymmetric. This means then
that the phenomenology~\eqref{eq:Z} is not applicable to the nodal
spectral functions of the cuprates.

\section{Dynes superconductors}
Very recently, a consistent two-lifetime phenomenology for
superconductors has been derived within the coherent potential
approximation, assuming a Lorentzian distribution of pair-breaking
fields and an arbitrary distribution of pair-conserving
disorder.\cite{Herman16} If we denote the pair-breaking and
pair-conserving scattering rates as $\Gamma$ and $\Gamma_s$,
respectively, then the result of Ref.~\onlinecite{Herman16} for the
Nambu-Gor'kov Green function of the disordered superconductor can be
written as
\begin{equation}
\hat{G}({\bf k},\omega)=
\frac{(1+i\Gamma_s/\Omega)\left[(\omega+i\Gamma)\tau_0
+{\bar \Delta}\tau_1\right]
+\varepsilon_{\bf k}\tau_3}
{(\Omega+i\Gamma_s)^2-\varepsilon_{\bf k}^2},
\label{eq:dynes_green}
\end{equation}
where 
\begin{equation}
\Omega(\omega)=\sqrt{(\omega+i\Gamma)^2-{\bar \Delta}^2}.
\label{eq:dynes_omega}
\end{equation}
Some useful properties of the function $\Omega(\omega)$ are described
in Appendix~A.  The tunneling density of states implied by the Green
function Eq.~\eqref{eq:dynes_green} is described by the Dynes formula
Eq.~\eqref{eq:dynes} with $\gamma=\Gamma$, and that is why
superconductors described by Eq.~\eqref{eq:dynes_green} have been
called Dynes superconductors in Ref.~\onlinecite{Herman16}. In this
work we shall keep this term.

It is worth pointing out that in absence of pair-breaking processes,
Eq.~\eqref{eq:dynes_green} reproduces the textbook results for
pair-conserving scattering, see
e.g. Refs.~\onlinecite{Marsiglio08,Zhu04}.  On the other hand, in the
opposite limit $\Gamma_s=0$ when only pair-breaking processes are
present, Eq.~\eqref{eq:dynes_green} coincides with the
phenomenology~\eqref{eq:Z} in the physically consistent case with
$\Gamma_0=\Gamma_1=\Gamma$. Moreover, in the normal state with
$\overline{\Delta}=0$ the Green function Eq.~\eqref{eq:dynes_green}
becomes diagonal and its matrix elements are
\begin{equation}
G_{11}(\mathbf{k},\omega)= 
\frac{1}{\omega - \varepsilon_{\mathbf{k}} + i\Gamma_n}, 
\quad
G_{22}(\mathbf{k},\omega)= 
\frac{1}{\omega + \varepsilon_{\mathbf{k}} + i\Gamma_n},
\label{eq:dynes_normal}
\end{equation}
where $\Gamma_n=\Gamma+\Gamma_s$ is the total scattering rate, which
involves both, pair-breaking as well as pair-conserving scattering
processes. Note that Eq.~\eqref{eq:dynes_normal} does not exhibit the
pathologies implied by Eq.~\eqref{eq:normal}.

One checks readily that the Green function Eq.~\eqref{eq:dynes_green}
is analytic in the upper half-plane of complex frequencies, as
required by causality, and that $\hat{G}({\bf k},\omega)\propto
\tau_0/|\omega|$ for $|\omega|\rightarrow \infty$.  In Appendix~B, we
present an explicit proof that Eq.~\eqref{eq:dynes_green} satisfies
the well-known sum rules for the zero-order moments of the electron
spectral function, in particular also the sum rule
Eq.~\eqref{eq:sumrule}. Moreover, in Appendix~D we prove that the
electron and hole spectral functions are positive-definite, as
required by general considerations.

In view of these observations, we believe that
Eq.~\eqref{eq:dynes_green} represents the simplest internally
consistent Green function for a superconductor with simultaneously
present pair-breaking and pair-conserving scattering processes. This
generic BCS-like Green function is parameterized by three energy
scales: scattering rates $\Gamma$ and $\Gamma_s$, as well as by the
gap parameter $\overline{\Delta}$. In what follows we present a
detailed analysis of its spectral properties.

\begin{figure*}[t]
\includegraphics[width = 17cm]{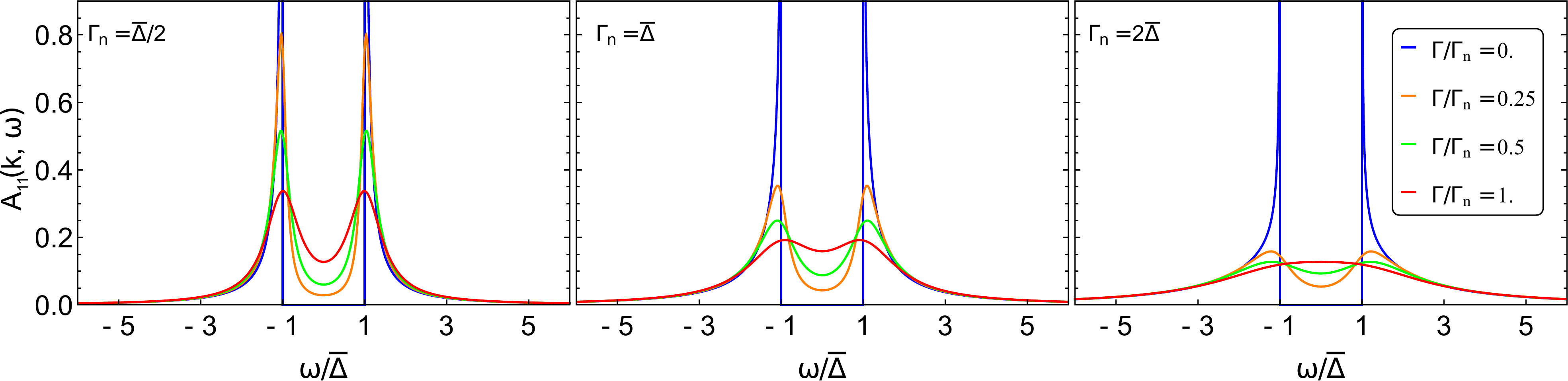}
\caption{Spectral functions $A_{11}(\mathbf{k},\omega)$ of the Dynes
  superconductor for an electron directly at the Fermi surface,
  $\varepsilon_{\bf k}=0$. The total scattering rate
  $\Gamma_n=\Gamma+\Gamma_s$ increases from the left to the right
  panel. The curves in each panel differ by the strength of the
  pair-breaking scattering rate $\Gamma$, while $\Gamma_n$ is kept
  fixed. The color coding is the same in all panels.}
\label{fig:arpes2}
\end{figure*}

Spectral functions of the Dynes superconductor for an electron with
momentum ${\bf k}$ fixed to lie inside the Fermi sea are shown in
Fig.~\ref{fig:arpes1}. The BCS quasiparticle peaks at
$\omega\approx\pm E_{\bf k}$ are seen to be broadened by the total
scattering rate $\Gamma_n$, irrespective of the ratio between
pair-breaking and pair-conserving scattering processes. The relative
importance of the two types of processes becomes important only in the
vicinity of the chemical potential. For $\Gamma=0$ a full spectral gap
appears for $|\omega|<\overline{\Delta}$, in agreement with the
Anderson theorem, and additional peaks appear in the spectral function
at $\omega=\pm \overline{\Delta}$. After switching on a finite pair
breaking rate $\Gamma\neq 0$, the spectral gap starts to fill in, and
at the same time the peaks at $\omega=\pm \overline{\Delta}$ get
smeared away. Finally, when $\Gamma=\Gamma_n$ and the pair-conserving
processes disappear completely, the spectral function is given by a
sum of two Lorentzians centered at $\omega=\pm E_{\bf k}$.

Spectral functions for an electron directly at the Fermi surface,
$\varepsilon_{\bf k}=0$, are somewhat different and they are shown in
Fig.~\ref{fig:arpes2}. The difference is caused by the fact that the
quasiparticle energies $\pm E_{\bf k}$ in this case coincide with $\pm
\overline{\Delta}$. Therefore only two peaks are present in the
spectral function, in contrast to the general case with four
peaks. However, the rest of the phenomenology can be simply related to
the case $\varepsilon_{\bf k}\neq 0$: the high-energy form of the spectral
functions is controlled exclusively by the total scattering rate
$\Gamma_n$, whereas finite pair-breaking fills in the spectral gap and
smears the peaks at $\omega=\pm \overline{\Delta}$.

In Appendix~C we complement the discussion of electron spectral
functions by studying the so-called momentum distribution
functions.\cite{Campuzano04} In addition to presenting explicit
formulas valid for any Eliashberg-type superconductor with only
frequency-dependent functions $Z(\omega)$ and $\phi(\omega)$, we also
show that making use of the momentum distribution functions, one can
determine the total scattering rate $\Gamma_n$ of a Dynes
superconductor in an alternative way.

In Fig.~\ref{fig:experiment} we demonstrate that
Eq.~\eqref{eq:dynes_green} can fit the experimentally observed
symmetrized spectral functions in the nodal region of the cuprates
with at least comparable quality as~Eq.~\eqref{eq:SelfEnergy}. The
number of fitting parameters is the same for both fits: two scattering
rates, the gap $\overline{\Delta}$, and the energy scale $\Lambda$
which determines the phenomenological background
$|\omega|/\Lambda^2$. This type of background description has been
used in all fits presented in Ref.~\onlinecite{Kondo15}.  We have
determined the fitting parameters by the standard least-squares
technique in the interval from -100~meV to 100~meV; their values are
$\overline{\Delta}=25$~meV, $\Gamma=3.7$~meV, $\Gamma_s=16$~meV, and
$\Lambda=103$~meV for the fit using Eq.~\eqref{eq:dynes_green}.  On
the other hand, we have found $\overline{\Delta}=27$~meV,
$\Gamma_0=0$~meV, $\Gamma_1=12$~meV, and $\Lambda=87$~meV for the fit
using Eq.~\eqref{eq:SelfEnergy}.

Both fits find roughly the same value of the gap $\overline{\Delta}$
and of the background parameter $\Lambda$, but the scattering rates
turn out to be quite different.  In view of the latter observation it
seems to be worthwhile to repeat the analysis of
Ref.~\onlinecite{Kondo15}, but with the ansatz
Eq.~\eqref{eq:dynes_green} for the electron Green function.  It
remains to be seen whether this type of analysis can be applied also
at temperatures above $T_c$, and what is the resulting temperature
dependence of the scattering rates $\Gamma$ and $\Gamma_s$.

The small value of the pair-breaking rate $\Gamma$ with respect to the
large pair-conserving rate $\Gamma_s$ implied by
Fig.~\ref{fig:experiment} is consistent with the observation that the
concept of the tomographic density of states is useful in the analysis
of the ARPES data.\cite{Reber12,Reber15} Since in an anisotropic
superconductor large-angle scattering is pair-breaking, the smallness
of $\Gamma$ implies that the dominant scattering processes (at least
in the nodal region and at low temperatures) have to be of the
forward-scattering type.

The importance of forward-scattering processes in the nodal region has
been confirmed recently also by an analysis of the momentum
distribution curves.\cite{Hong14} As for the microscopic origin of the
forward scattering, it has been argued that it can be caused by
elastic scattering on disorder located outside the CuO$_2$
planes.\cite{Abrahams00} Other explanations include scattering on
(quasi) static long-range fluctuations, perhaps due to competing
order, or scattering on low-energy long-wavelength emergent gauge
fields.\cite{Lee06} These different scenaria can be distinguished by
different dependence on temperature and/or Fermi-surface location and
further experimental work is needed to discriminate between them.

\section{Conclusions}
To summarize, we have shown that the phenomenological
self-energy~Eq.~\eqref{eq:SelfEnergy}, which has been proposed
theoretically in Refs.~\onlinecite{Norman98,Chubukov07} and applied
recently in Ref.~\onlinecite{Kondo15}, is internally consistent only
in the case when $\Gamma_0=\Gamma_1$; in this case the electron
spectral function in the superconducting state is a sum of two
Lorentzians.

\begin{figure}[t]
\includegraphics[width = 8cm]{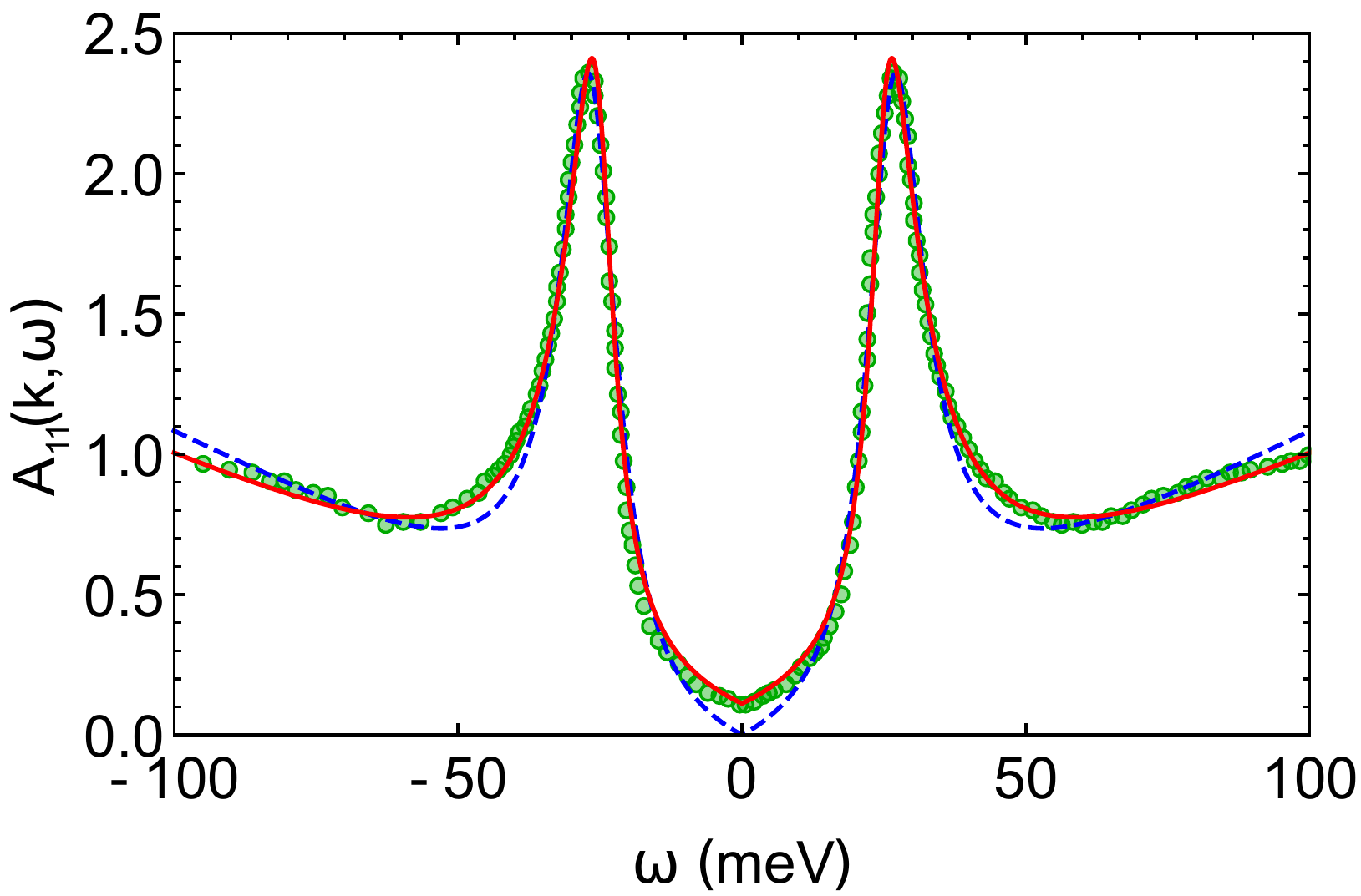}
\caption{Experimentally observed low-temperature symmetrized spectral
  functions at the Fermi level reported in Ref.~\onlinecite{Kondo15}
  for optimally doped Bi2212 at angle $\phi=24^\circ$. Also shown are
  least-square fits in the region from -100~meV to +100~meV around the
  Fermi level which make use of Eq.~\eqref{eq:dynes_green} (red solid
  line) and of Eq.~\eqref{eq:SelfEnergy} (blue dashed line). The
  values of the fitting parameters are shown in the main text.}
\label{fig:experiment}
\end{figure}

The simplest consistent genuine two-lifetime Green function of a
superconductor is given by Eq.~\eqref{eq:dynes_green}. This model
depends on two scattering rates: the pair-breaking scattering rate
$\Gamma$ and the pair-conserving scattering rate $\Gamma_s$. The Green
function Eq.~\eqref{eq:dynes_green} implies that the density of states
is described by the Dynes formula Eq.~\eqref{eq:dynes} with
$\gamma=\Gamma$ and the electron spectral functions exhibit more
structure than might be expected naively, see
Figs.~\ref{fig:arpes1},\ref{fig:arpes2}.

The Green function Eq.~\eqref{eq:dynes_green} is analytic in the upper
half-plane, it has the correct large-frequency asymptotics, its
diagonal spectral functions are positive-definite, and it satisfies
the exact sum rules Eq.~\eqref{eq:sumrule} and
Eq.~\eqref{eq:sum_rule_1}. Moreover, in the three limiting cases of
either $\Gamma=0$, or $\Gamma_s=0$, or $\overline{\Delta}=0$, it
reduces to the well-known results. Therefore, although
Eq.~\eqref{eq:dynes_green} has been originally derived only for a
special distribution of pair-breaking fields within the coherent
potential approximation, we believe that it represents a {\it generic}
two-lifetime Green function of a superconductor.

Our results provide a (in principle) straightforward recipe for
extracting the scattering rates $\Gamma$ and
$\Gamma_n=\Gamma+\Gamma_s$ from experimental data: the pair-breaking
scattering rate $\Gamma$ is best determined from the tunneling (or, in
anisotropic superconductors, tomographic\cite{Reber12}) density of
states, whereas the total scattering rate $\Gamma_n$ may be extracted
from the widths of the quasiparticle peaks in spectral functions, see
Fig.~\ref{fig:arpes1}. Alternatively, as shown in Appendix~C, the
scattering rate $\Gamma_n$ can be determined from the width of the
momentum distribution functions, and it enters also the analysis of
optical conductivity.\cite{Herman17}

Obviously, description of superconductors making use of
Eq.~\eqref{eq:dynes_green} can be quantitatively correct only at
energies smaller than the typical boson energies of the studied
system. At higher energies, application of a full-fledged
Eliashberg-type theory\cite{Marsiglio08} - but extended so as to allow
for processes leading to Eq.~\eqref{eq:dynes_green} at low energies -
is unavoidable. For completeness, in Appendix~C we
have described a procedure which, starting from the assumption of only
frequency-dependent Eliashberg functions $Z(\omega)$ and
$\Delta(\omega)$, allows for their complete determination from ARPES
data by combining two approaches: the momentum distribution technique
and the tomographic density of states.

Finally, in Fig.~\ref{fig:experiment} we have demonstrated that the
low-temperature ARPES data in the nodal region of the cuprates can be
fitted well using Eq.~\eqref{eq:dynes_green}. Our results confirm
previous claims about the importance of forward-scattering processes
in this region, but identification of their physical origin will
require further detailed angle- and temperature-dependent studies.

\appendix
\section{Properties of the function $\Omega(\omega)$}
Let us decompose the function $\Omega(\omega)$ defined by
Eq.~\eqref{eq:dynes_omega} into its real and imaginary parts,
$\Omega=\Omega_1+i\Omega_2$.  One finds readily that $\Omega_{1,2}$
should satisfy the relations
\begin{equation}
\Omega_1\Omega_2=\omega \Gamma,
\qquad
\Omega_1^2-\Omega_2^2=\nu^2,
\label{eq:omega12}
\end{equation}
where $\nu^2=\omega^2-\overline{\Delta}^2-\Gamma^2$.  Our sign
convention leads then to the following explicit expressions for
$\Omega_{1,2}$:
\begin{eqnarray*}
\Omega_1(\omega)&=&{\rm sgn}(\omega)
\sqrt{\left[\sqrt{\nu^4+4\omega^2\Gamma^2}+\nu^2\right]/2},
\\
\Omega_2(\omega)&=&
\sqrt{\left[\sqrt{\nu^4+4\omega^2\Gamma^2}-\nu^2\right]/2}.
\end{eqnarray*}
Note that $\Omega_1(\omega)$ is an odd function of $\omega$, while
$\Omega_2(\omega)$ is positive definite and even.  A straightforward
calculation shows that for $\omega>0$ the following inequalities are
valid:
\begin{equation}
\Omega_1\leq \omega,
\qquad
\Omega_2\geq \Gamma.
\label{eq:inequalities_2}
\end{equation}
These inequalities will be used in Appendix~D.

It is worth pointing out that the function $\Omega_1(\omega)$
characterizing the Dynes superconductor is in principle directly
measurable in low-temperature tunneling experiments. In fact, it is
well known that in such experiments the derivative of the
current-voltage characteristics, $dI/dV$, is proportional to the
tunneling density of states $N(\omega)$ with $\omega=eV$. But, since
$N(\omega)\propto d\Omega_1/d\omega$, the function $\Omega_1(\omega)$
is proportional to the measured function $I=I(V)$.

\section{Sum rules for the Dynes superconductors}
In this Appendix we prove that Eq.~\eqref{eq:dynes_green} satisfies
the sum rules for the zero-order moments of the electron spectral
function.  To this end, let us introduce an auxiliary complex function
$F(\omega)$ of the real frequency $\omega$,
$$
F(\omega)=\varepsilon_{\bf k}^2-
\left[\Omega(\omega)+i\Gamma_s\right]^2.
$$ 
Note that the function $F(\omega)$ depends also on the momentum ${\bf
  k}$, but for the sake of simplicity this dependence will not be
displayed explicitly.

\begin{figure}[t]
\includegraphics[width = 8cm]{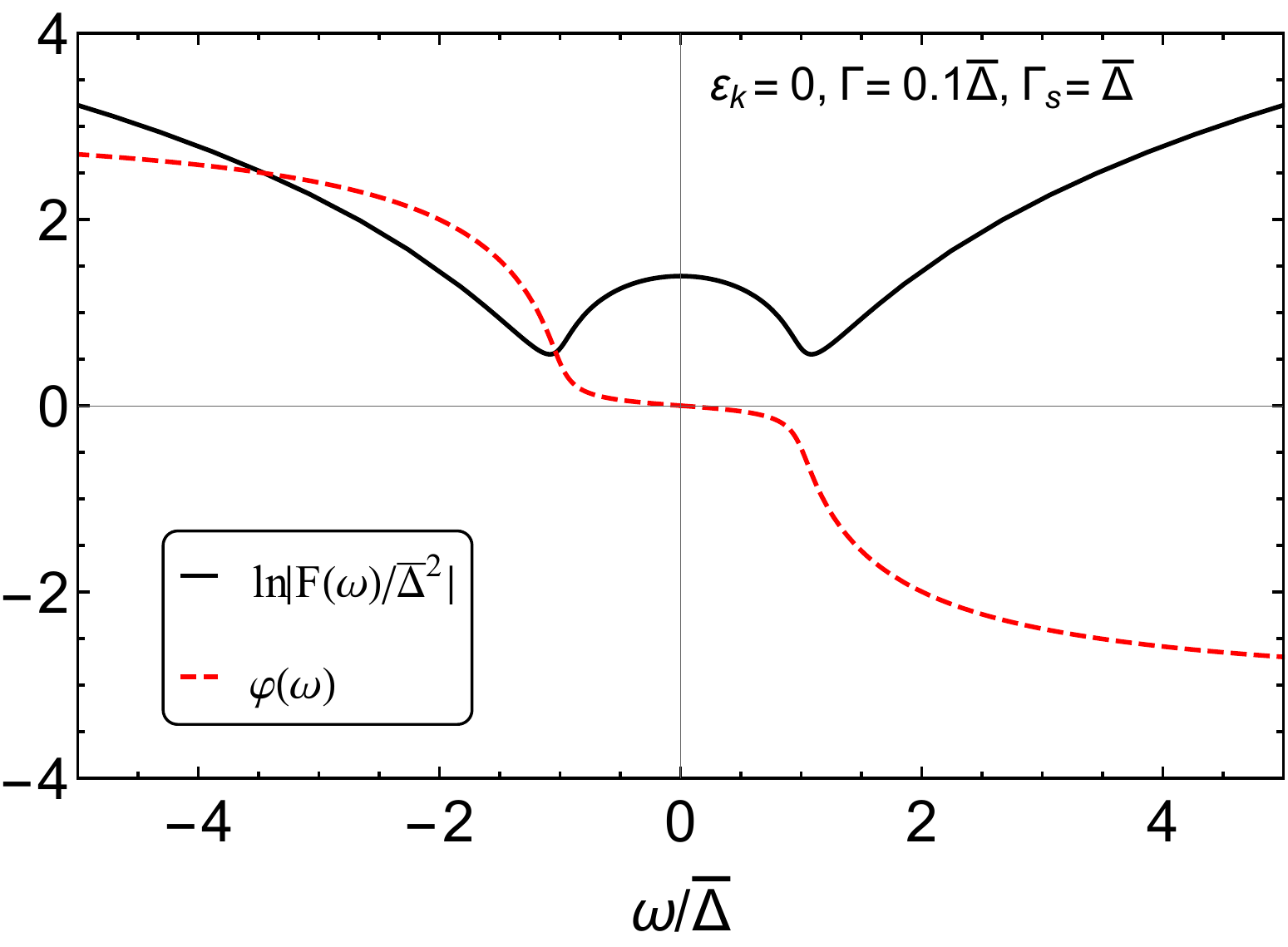}
\caption{Real and imaginary parts of the function $H(\omega)$.}
\label{fig:function_f}
\end{figure}

Let us furthermore define the function
$$
H(\omega)=\ln F(\omega)=\ln |F(\omega)|+i\varphi(\omega).
$$ 
In the second equality we have represented the complex function
$F(\omega)=|F(\omega)|\exp\{i\varphi(\omega)\}$ in terms of its
amplitude $|F(\omega)|$ and phase $\varphi(\omega)$ constrained to the
interval $(-\pi,\pi)$.  A plot of the real and imaginary parts of the
function $H(\omega)$ is shown in Fig.~\ref{fig:function_f}.  Note that
the phase $\varphi(\omega)$ is an odd function of frequency and its
asymptotic values are $\varphi(\pm\infty)=\mp \pi$.

Making use of the function $H(\omega)$, the Nambu-Gor'kov Green
function~\eqref{eq:dynes_green} can be written in the following
elegant form:
\begin{equation}
\hat{G}({\bf k},\omega)=
\frac{1}{2}\left[\frac{\partial H}{\partial\omega}\tau_0
-\frac{\partial H}{\partial\overline{\Delta}}\tau_1
-\frac{\partial H}{\partial\varepsilon_{\bf k}^{}}\tau_3\right].
\label{eq:dynes_elegant}
\end{equation}
From here follows the following explicit expression for the
Nambu-Gor'kov spectral function, defined as usual by
$\hat{A}({\bf k},\omega)=
-\pi^{-1}{\rm Im}\hat{G}({\bf k},\omega)$:
\begin{equation}
\hat{A}({\bf k},\omega)=
\frac{1}{2\pi}\left[-\frac{\partial \varphi}{\partial\omega}\tau_0
+\frac{\partial \varphi}{\partial\overline{\Delta}}\tau_1
+\frac{\partial \varphi}{\partial\varepsilon_{\bf k}^{}}\tau_3\right].
\label{eq:dynes_a_elegant}
\end{equation}
Equation~\eqref{eq:dynes_a_elegant} forms the starting point of our
discussion of the sum rules. 

Using the oddness of the function $\varphi(\omega)$ and of its
asymptotic values, one finds readily that
Eq.~\eqref{eq:dynes_a_elegant} implies the matrix equation
\begin{equation}
\int_{-\infty}^\infty d\omega \hat{A}({\bf k},\omega)=\tau_0,
\label{eq:sum_rule_1}
\end{equation}
in perfect agreement with the well-known exact sum rule for the
zero-order moment of the spectral function.

Next we prove that Eq.~\eqref{eq:dynes_a_elegant} satisfies the exact
sum rule Eq.~\eqref{eq:sumrule}. Since ${\rm Tr}\hat{A}({\bf
  k},\omega)= -\pi^{-1}\partial \varphi/\partial\omega$, we have to
prove that
$$
\int_{-\infty}^{\infty}\frac{d\omega}{1+e^{-\omega/T}}
\frac{\partial \varphi}{\partial\omega}=-\pi.
$$
By calculating the integral on the left-hand side by parts, 
our task reduces to proving the equality
$$
\int_{-\infty}^{\infty}\frac{d\omega}{4T \cosh^2(\omega/2T)}
\varphi(\omega)=0.
$$
But, since $\varphi(\omega)$ is odd, this last equality is trivially
satisfied. Thus we have proven that the Dynes superconductors satisfy
Eq.~\eqref{eq:sumrule}. 

For the sake of completeness, let us note that the full matrix form of
the sum rule Eq.~\eqref{eq:sumrule} reads
\begin{equation}
\int_{-\infty}^\infty \frac{d\omega}{1+e^{-\omega/T}} 
\hat{A}({\bf k},\omega)=
\left(\begin{array}{cc}
1-n_{\bf k} & b_{\bf k}
\\
b_{\bf k} & n_{\bf k}
\end{array}
\right),
\label{eq:sum_rule_2}
\end{equation}
where $n_{\bf k}=n_{{\bf k}\uparrow}=n_{-{\bf k}\downarrow}$, $b_{\bf
  k}=\langle c^{}_{{\bf k}\uparrow}c^{}_{-{\bf k}\downarrow}\rangle
=\langle c^\dagger_{-{\bf k}\downarrow}c^\dagger_{{\bf
    k}\uparrow}\rangle$, and the thermodynamic expectation values
$n_{\bf k}$ and $b_{\bf k}$ are given by
\begin{eqnarray*}
n_{\bf k}&=&\frac{1}{2}-\int_0^\infty \frac{d\omega}{2\pi}
\frac{\partial \varphi}{\partial\varepsilon^{}_{\bf k}}
\tanh\frac{\omega}{2T},
\\
b_{\bf k}&=&\int_0^\infty \frac{d\omega}{2\pi}
\frac{\partial \varphi}{\partial\overline{\Delta}}
\tanh\frac{\omega}{2T}.
\end{eqnarray*}

It should be pointed out that sum rules for higher-order moments of
the spectral function which generalize
Eqs.~(\ref{eq:sum_rule_1},\ref{eq:sum_rule_2}) can be also derived,
but their right-hand sides depend on the Hamiltonian of the problem.
Such sum rules therefore do not provide useful checks of the
phenomenological Green function Eq.~\eqref{eq:dynes_green}.

\section{Momentum distribution functions in the Eliashberg theory}
We have already noted that, within the Eliashberg theory, the
functions $Z(\omega)$ and $\phi(\omega)$ usually depend only on
frequency $\omega$ and are independent of the momentum ${\bf k}$.
Quite some time ago, it has been pointed out that in such cases it is
useful to study the spectral function $A_{11}({\bf k},\omega)$ for
fixed frequency $\omega$ as a function of the bare electron energy
$\varepsilon_{\bf k}$, the so-called momentum distribution
function.\cite{Campuzano04} To simplify the formulas, in this Appendix
we will replace $A_{11}({\bf k},\omega)$ by $A(\varepsilon,\omega)$.

\begin{figure}[t]
\includegraphics[width = 8cm]{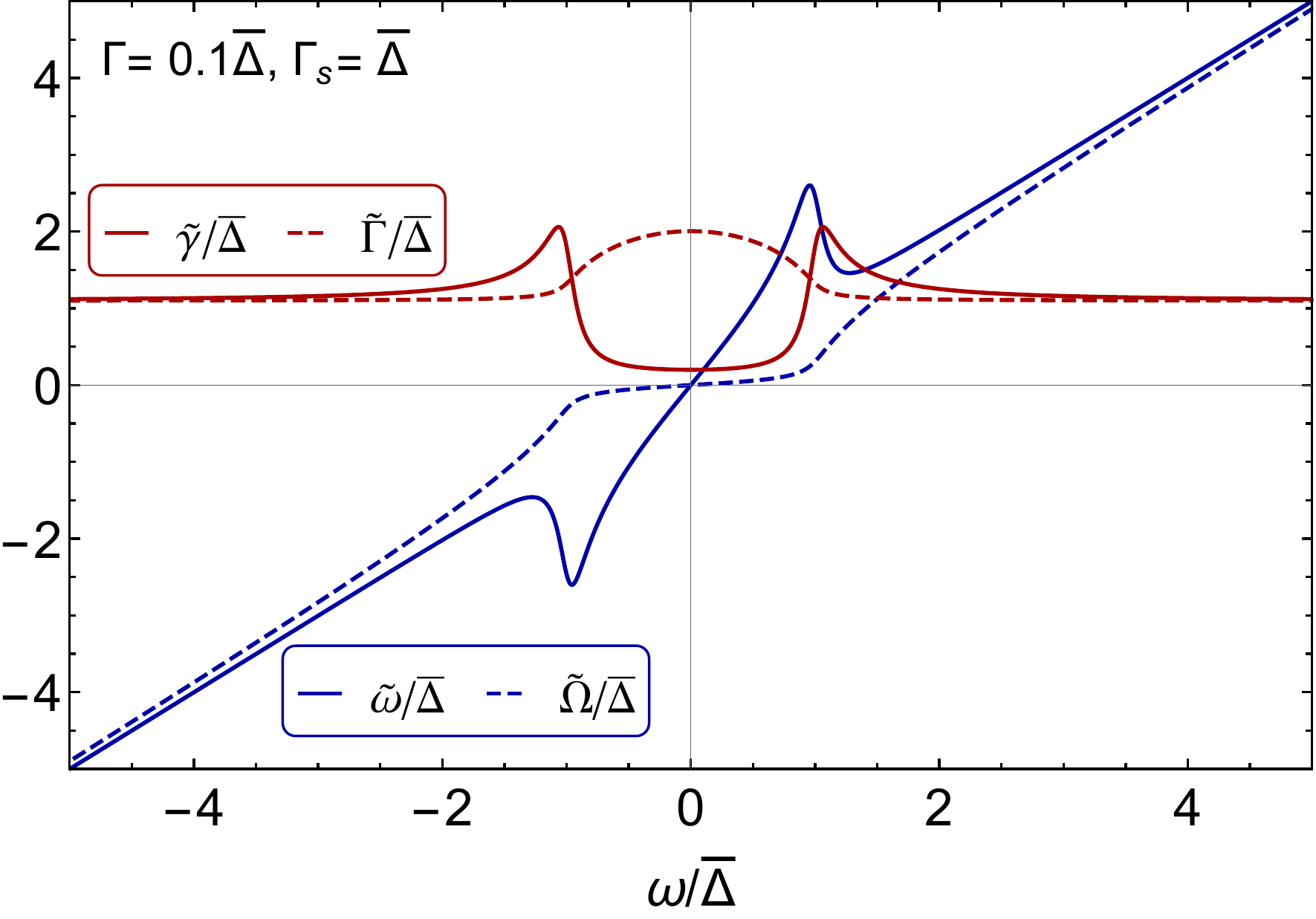}
\caption{Functions $\widetilde{\omega}(\omega)$,
  $\widetilde{\gamma}(\omega)$, $\widetilde{\Omega}(\omega)$, and
  $\widetilde{\Gamma}(\omega)$ for a Dynes superconductor.}
\label{fig:tildes}
\end{figure}

Instead of the two complex functions $Z(\omega)$ and $\phi(\omega)$,
let us introduce the following four real functions of frequency
$\widetilde{\omega}(\omega)$, $\widetilde{\gamma}(\omega)$,
$\widetilde{\Omega}(\omega)$, and $\widetilde{\Gamma}(\omega)$:
\begin{eqnarray}
\omega Z&=&\widetilde{\omega}+i\widetilde{\gamma},
\nonumber
\\
\sqrt{(\omega Z)^2-\phi^2}&=&\widetilde{\Omega}+i\widetilde{\Gamma}.
\label{eq:tildes}
\end{eqnarray}
To illustrate their symmetries and typical form, in
Fig.~\ref{fig:tildes} we plot the functions
$\widetilde{\omega}(\omega)$, $\widetilde{\gamma}(\omega)$,
$\widetilde{\Omega}(\omega)$, and $\widetilde{\Gamma}(\omega)$ for a
Dynes superconductor.

After a tedious but straightforward calculation the spectral function
of a general Eliashberg superconductor can be written as
\begin{eqnarray}
A(\varepsilon,\omega)&=&
\frac{1}{2}
\left[\frac{\widetilde{\gamma}}{\widetilde{\Gamma}}+1\right]
\delta_{\widetilde{\Gamma}}(\varepsilon-\widetilde{\Omega})
+\frac{1}{2}
\left[\frac{\widetilde{\gamma}}{\widetilde{\Gamma}}-1\right]
\delta_{\widetilde{\Gamma}}(\varepsilon+\widetilde{\Omega})
\nonumber
\\
&+&
\frac{1}{2}
\left[\frac{\widetilde{\omega}}{\widetilde{\Omega}}
-\frac{\widetilde{\gamma}}{\widetilde{\Gamma}}\right]
\frac{4\pi\widetilde{\Omega}^2}{\widetilde{\Gamma}}
\delta_{\widetilde{\Gamma}}(\varepsilon-\widetilde{\Omega})
\delta_{\widetilde{\Gamma}}(\varepsilon+\widetilde{\Omega}),
\label{eq:momentum}
\end{eqnarray}
where we have introduced the notation
$$
\delta_{\widetilde{\Gamma}}(x)=\frac{1}{\pi}
\frac{\widetilde{\Gamma}}{x^2+\widetilde{\Gamma}^2}
$$ 
for a Lorentzian with width $\widetilde{\Gamma}$.  According to
Eq.~\eqref{eq:momentum}, the spectral function
$A(\varepsilon,\omega)$, when viewed as a function of energy
$\varepsilon$ at fixed frequency $\omega$, consists of three
terms. The first two terms are Lorentzians, whereas the third term is
a product of two Lorentzians.

When the measured momentum distribution functions are fitted by
Eq.~\eqref{eq:momentum}, $\widetilde{\Omega}$ can be determined from
the positions of the Lorentzians and $\widetilde{\Gamma}$ is given by
their widths. Finally, from the relative weights of the three terms in
Eq.~\eqref{eq:momentum} one can determine the ratios
$\widetilde{\gamma}/\widetilde{\Gamma}$ and
$\widetilde{\omega}/\widetilde{\Omega}$. With all four functions
$\widetilde{\omega}(\omega)$, $\widetilde{\gamma}(\omega)$,
$\widetilde{\Omega}(\omega)$, and $\widetilde{\Gamma}(\omega)$ known,
one obtains full information about the superconducting state.  This
idea has been used in an impressive set of recent papers, see
Ref.\onlinecite{Bok16} and references therein.

One should note, however, that in order to determine all four
parameters $\widetilde{\omega}$, $\widetilde{\gamma}$,
$\widetilde{\Omega}$, and $\widetilde{\Gamma}$, it is necessary to
resolve all three terms in Eq.~\eqref{eq:momentum}, together with
their relative weights. But at sufficiently large frequencies we
should expect that $\widetilde{\Gamma}\ll |\widetilde{\Omega}|$, see
Fig.~\ref{fig:tildes}. In this case the following approximate equality
is valid
$$
\frac{4\pi\widetilde{\Omega}^2}{\widetilde{\Gamma}}
\delta_{\widetilde{\Gamma}}(\varepsilon-\widetilde{\Omega})
\delta_{\widetilde{\Gamma}}(\varepsilon+\widetilde{\Omega})
\approx 
\delta_{\widetilde{\Gamma}}(\varepsilon-\widetilde{\Omega})
+\delta_{\widetilde{\Gamma}}(\varepsilon+\widetilde{\Omega}),
$$ 
which means that the product of two Lorentzians can not be
distinguished from their sum.  Inserting this equality into
Eq.~\eqref{eq:momentum}, one finds readily that the spectral function
$A(\varepsilon,\omega)$ is given by a sum of only two Lorentzians,
\begin{eqnarray*}
A(\varepsilon,\omega)\approx
\frac{1}{2}
\left[\frac{\widetilde{\omega}}{\widetilde{\Omega}}+1\right]
\delta_{\widetilde{\Gamma}}(\varepsilon-\widetilde{\Omega})
+\frac{1}{2}
\left[\frac{\widetilde{\omega}}{\widetilde{\Omega}}-1\right]
\delta_{\widetilde{\Gamma}}(\varepsilon+\widetilde{\Omega}).
\end{eqnarray*}
But if this is the case, then from fits to the momentum distribution
function one can determine only $\widetilde{\Omega}$,
$\widetilde{\Gamma}$, and $\widetilde{\omega}$, but not
$\widetilde{\gamma}$. In other words, we do not have access to the
pairing function
$\phi^2(\omega)=(\widetilde{\omega}+i\widetilde{\gamma})^2
-(\widetilde{\Omega}+i\widetilde{\Gamma})^2$ in this frequency limit.

There is yet another reason why fits to the momentum distribution
function can provide reliable estimates of the Eliashberg parameters
only for $|\omega|\lesssim \overline{\Delta}$: namely, this technique
requires that both ratios, $\widetilde{\gamma}/\widetilde{\Gamma}$ and
$\widetilde{\omega}/\widetilde{\Omega}$, are sufficiently different
from 1, so that the weights of the second and third terms in
Eq.~\eqref{eq:momentum} can be determined precisely. But
Fig.~\ref{fig:tildes} clearly shows that this criterion is satisfied
only for $|\omega|\lesssim \overline{\Delta}$.

Does this mean that the Eliashberg problem of finding the functions
$Z(\omega)$ and $\Delta(\omega)$ can not be solved in the frequency
range $\overline{\Delta}\lesssim |\omega|$? The answer is no: it has
been pointed out recently\cite{Bzdusek15,Bok16} that, by applying the
powerful inversion technique developed in Ref.~\onlinecite{Galkin74},
it is possible to extract the complex gap function
$\Delta(\omega)=\phi(\omega)/Z(\omega)$ from the measured tomographic
density of states. When this knowledge is combined with the momentum
distribution technique - which allows for a relatively straightforward
determination of $\widetilde{\Omega}$ and $\widetilde{\Gamma}$ in the
limit $\overline{\Delta}\lesssim |\omega|$ with one Lorentzian only -
making use of the expression
$$
Z(\omega)=\frac{\widetilde{\Omega}+i\widetilde{\Gamma}}
{\sqrt{\omega^2-\Delta^2(\omega)}}
$$
one can determine also the second Eliashberg function $Z(\omega)$, thereby
solving the Eliashberg problem.\cite{Bok16}

Finally, let us note that the momentum distribution functions can be
useful also in the special case of the Dynes superconductors described
by Eq.~\eqref{eq:dynes_green}.  In fact, since in the frequency range
$\overline{\Delta}\lesssim |\omega|$ the width of the observable
Lorentzian in the momentum distribution function of a Dynes
superconductor is $\widetilde{\Gamma}\approx\Gamma_n$, this gives us
an independent procedure for measuring the total scattering rate
$\Gamma_n=\Gamma+\Gamma_s$.

\section{Proof of the inequalities $A_{ii}({\bf k},\omega)\geq 0$ 
for the Dynes superconductors}
In this Appendix we will prove that the diagonal spectral functions
$A_{ii}({\bf k},\omega)$ of the Dynes superconductors are
positive-definite, as required by general considerations.

To this end, let us first note that the diagonal components of the
Nambu-Gor'kov Green function within the Eliashberg theory read as
$$
G_{ii}({\bf k},\omega)=
\frac{\widetilde{\omega}+i\widetilde{\gamma}\pm\varepsilon_{\bf k}}
{(\widetilde{\Omega}+i\widetilde{\Gamma})^2-\varepsilon_{\bf k}^2}.
$$
Since $A_{ii}({\bf k},\omega)=-\pi^{-1}{\rm Im}G_{ii}({\bf k},\omega)$,
from here it follows that the requirement 
$A_{ii}({\bf k},\omega)\geq 0$ is equivalent to 
$$
2\widetilde{\Omega}\widetilde{\Gamma}\widetilde{\omega}
-\widetilde{\gamma}(\widetilde{\Omega}^2-\widetilde{\Gamma}^2)
\geq
-\widetilde{\gamma}\varepsilon_{\bf k}^2
\mp 
2\widetilde{\Omega}\widetilde{\Gamma}\varepsilon_{\bf k},
$$
which has to hold for all $\varepsilon_{\bf k}$ and $\omega$.
Maximizing the expression on the right-hand side with respect to
$\varepsilon_{\bf k}$, this requirement can be rewritten as
$$
\frac{\widetilde{\gamma}^2\widetilde{\Gamma}^2}
{\widetilde{\gamma}^2+\widetilde{\Gamma}^2}
+
\widetilde{\omega}\widetilde{\Omega}
\frac{2\widetilde{\gamma}\widetilde{\Gamma}}
{\widetilde{\gamma}^2+\widetilde{\Gamma}^2}
\geq
\widetilde{\Omega}^2,
$$ 
which has to be valid for all frequencies $\omega$.  Since the
first term on the left-hand side is obviously positive, it follows
that it is sufficient to show that
$$
\frac{\widetilde{\omega}}{\widetilde{\Omega}}
\geq
\frac{1}{2}\left(
\frac{\widetilde{\gamma}}{\widetilde{\Gamma}}
+\frac{\widetilde{\Gamma}}{\widetilde{\gamma}}
\right).
$$ 
In order to prove this latter inequality, we will prove the
following two simpler inequalities:
\begin{equation}
\frac{\widetilde{\omega}}{\widetilde{\Omega}}
\geq
\frac{\widetilde{\gamma}}{\widetilde{\Gamma}},
\qquad
\frac{\widetilde{\omega}}{\widetilde{\Omega}}
\geq
\frac{\widetilde{\Gamma}}{\widetilde{\gamma}}.
\label{eq:inequalities}
\end{equation}
In view of the symmetries illustrated by Fig.~\ref{fig:tildes}, one
checks easily that it is sufficient to prove that these inequalities
hold for $\omega>0$.

So far, our discussion was valid for any Eliashberg superconductor.
Now we specialize to the Dynes superconductors. Making use of
Eqs.~\eqref{eq:tildes},\eqref{eq:dynes_green} one finds easily that in
this case the quantities $\widetilde{\omega}$, $\widetilde{\Omega}$,
$\widetilde{\Gamma}$, and $\widetilde{\gamma}$ can be written in terms
of the functions $\Omega_1$ and $\Omega_2$ introduced in Appendix~A as
\begin{eqnarray*}
\widetilde{\Omega}&=&\Omega_1,
\\
\widetilde{\Gamma}&=&\Omega_2+\Gamma_s,
\\
\widetilde{\omega}&=&\omega+
\Gamma_s\frac{\omega\Omega_2-\Gamma\Omega_1}
{\Omega_1^2+\Omega_2^2},
\\
\widetilde{\gamma}&=&\Gamma+
\Gamma_s\frac{\omega\Omega_1+\Gamma\Omega_2}
{\Omega_1^2+\Omega_2^2}.
\end{eqnarray*}
Let us note in passing that these expressions justify the results
plotted in Fig.~\ref{fig:tildes}.  

Next we plug the expressions for $\widetilde{\omega}$,
$\widetilde{\Omega}$, $\widetilde{\Gamma}$, and $\widetilde{\gamma}$
into Eqs.~\eqref{eq:inequalities}. If one makes use of the equalities
Eqs.~\eqref{eq:omega12} and of the inequalities
Eqs.~\eqref{eq:inequalities_2}, after some straightforward algebra one
can check that the inequalities Eqs.~\eqref{eq:inequalities} are
satisfied. This completes the proof that, for the Dynes
superconductors, the inequalities $A_{ii}({\bf k},\omega)\geq 0$ are
valid.

\begin{acknowledgments}
This work was supported by the Slovak Research and Development Agency
under contracts No.~APVV-0605-14 and No.~APVV-15-0496, and by the
Agency VEGA under contract No.~1/0904/15.
\end{acknowledgments}

\end{document}